\documentstyle[aasms4,12pt,psfig]{article}

\def\hb{\hfill\break}

\def\ale{\mathrel{\hbox{\rlap{\hbox{\lower4pt\hbox{$\sim$}}}\hbox{$<$}}}}
\def\age{\mathrel{\hbox{\rlap{\hbox{\lower4pt\hbox{$\sim$}}}\hbox{$>$}}}}

\def\gsim{\mathrel{\hbox{\rlap{\lower.55ex \hbox {$\sim$}}
                   \kern-.3em \raise.4ex \hbox{$>$}}}}
\def\lsim{\mathrel{\hbox{\rlap{\lower.55ex \hbox {$\sim$}}
                   \kern-.3em \raise.4ex \hbox{$<$}}}}

\def\Sref#1{\S\ref{#1}}

\def\grb{GRB\thinspace{990123}}

\makeatletter
\renewenvironment{deluxetable}[1]{\def\pt@format{\string#1}%
\set@tblnotetext\global\pt@ncol=0\global\pt@column=0\global\pt@page=1%
\def\pt@addcol{\global\advance\pt@ncol by\@ne}}%
{\pt@width\wd\pt@box\box\pt@box\vskip-0.5cm\spew@ptblnotes%
\typeout{Page \the\pt@page\space of table \thetable\space has been set
to
width \the\pt@width\space with \the\pt@nlines\space lines per page}%
\endcenter\vskip-0.5cm\end@float}
\def\startdata{\pt@line=0\pt@calcnlines%
\ifdim\pt@width>\z@\def\@halignto{to \pt@width}\else\def\@halignto{}\fi%
\let\fnum@table=\fnum@ptable\set@mkcaption%
\@float{table}[t]\center\caption{\pt@caption}\leavevmode%
\setbox\pt@box=\pt@tabular{\pt@format}\pt@head}
\def\thebibliography{\subsection*{REFERENCES}
\list{}{\labelwidth3em\leftmargin\labelwidth\labelsep\z@\parsep\z@
\itemsep\z@\itemindent-3em\usecounter{enumi}}
\def\refpar{\relax}
\def\newblock{\hskip .11em plus .33em minus .07em}
\sloppy\clubpenalty4000\widowpenalty4000
\sfcode`\.=1000\relax}
\makeatother

\lefthead{Kulkarni et al.}
\righthead{Radio Flare from GRB 990123}
\begin{document}
 
 
\received{30 March 1999}
\accepted{15 July 1999}
\title{\large \bf Discovery of a Radio Flare from GRB 990123}

\author
{
  S.~R.~Kulkarni\altaffilmark{1},
  D.~A.~Frail\altaffilmark{2},
  R.~Sari\altaffilmark{3},
  G.~H.~Moriarty-Schieven\altaffilmark{4,5},
  D.~S.~Shepherd\altaffilmark{1},
  P.~Udomprasert\altaffilmark{1},
  A.~C.~S.~Readhead\altaffilmark{1},
  J.~S.~Bloom\altaffilmark{1},
  M.~Feroci\altaffilmark{6} \&\
  E.~Costa\altaffilmark{6} 
}

\altaffiltext{1}{California Institute of Technology, 
Owens Valley Radio Observatory
105-24, Pasadena, CA 91125, USA}
 
\altaffiltext{2}
{National Radio Astronomy Observatory, P.~O.~Box O,
 Socorro, NM 87801, USA}

\altaffiltext{3}
{California Institute of Technology,
 Theoretical Astrophysics 
 103-33, Pasadena, CA 91125, USA} 

\altaffiltext{4}
{Joint Astronomy Centre, 660 N. A'ohoku Place,
 Hilo, HI 96720}

\altaffiltext{5} {Herzberg Institute of Astrophysics, 5071 West Saanich
Road, Victoria, BC, V8X 4M6, Canada}

\altaffiltext{6}
{Istituto di Astrofisica Spaziale, CNR,
via Fosso del Cavaliere, Roma I-00133, Italy}

\begin{abstract}
We report the discovery of a radio counterpart to GRB~990123. In contrast
to previous well-studied radio afterglows which rise to peak flux on
a timescale of a week and then decay over several weeks to months, the
radio emission from this GRB was clearly detected one day after the burst,
after which it rapidly faded away. The simplest interpretation of this
``radio flare'' is that it arises from the reverse shock.  In the framework
of the afterglow models discussed to date, a forward shock origin for
the flare is ruled out by our data.  However, at late times,
some radio afterglow emission (commensurate with the observed late-time
optical emission, the optical afterglow) is expected from the forward
shock. The relative faintness of the observed late-time radio emission
provides an independent indication for a jet-like geometry in this GRB.
We use the same radio observations to constrain two key parameters of
the forward shock, peak flux and peak frequency, to within a factor of
two. These values are inconsistent with the notion advocated by several
authors that the prompt optical emission detected by ROTSE smoothly joins
the optical afterglow emission.   Finally, with hindsight we now recognize
another such radio flare and this suggests that one out of eight GRBs
has a detectable radio flare. This abundance coupled with the reverse
shock interpretation suggests that the radio flare phenomenon has the
potential to shed new light into the physics of reverse shocks in GRBs.

\end{abstract}

\keywords{gamma rays: bursts; radio continuum: general; shock waves;}
 
\section{Introduction \label{sec:introduction}}

BeppoSAX ushered in 1999 with the discovery of a super-bright $\gamma$-ray
burst, \grb. This GRB was intensively studied by many groups worldwide.
The burst is notable for the richness of new results: the discovery of
prompt optical emission by ROTSE (Akerlof et al. 1999), the discovery
of the brightest optical afterglow to date and its redshift (Kulkarni
et al. 1999), and a break in the optical afterglow light curve (Kulkarni
et al. 1999, Fruchter et al. 1999, Castro-Tirado et al. 1999).

To date, as a result of our radio afterglow program at the Very Large
Array (VLA)\footnotemark\footnotetext{The VLA is operated by the
  National Radio Astronomy Observatory which is a facility of the
  National Science Foundation operated under a cooperative agreement
  by Associated Universities, Inc.} and the Australia Telescope
Compact Array (ATCA) we have detected radio afterglow emission in nine
GRBs (e.g. Frail et al.  1997b; Taylor et al. 1998). Observations of
radio afterglow have resulted in the first direct proof for
relativistic expansion in GRBs (Frail et al. 1997b) and robust
estimates of the total burst energy (Frail, Waxman \&\ Kulkarni 1999).

The radio afterglow, when detected, exhibit the following pattern: the
flux rises gradually to a maximum,  persists for weeks to 
months before fading away.  It is against this backdrop that we report,
in this {\it Letter}, the discovery of a new phenomenon in the radio
afterglow of GRBs -- the discovery of a short lived ($\sim$30 hours)
radio emission in GRB 990123.  This discovery, in a manner similar to the ROTSE
discovery, adds yet another diagnostic -- a probe of the reverse shock --
to the study of GRBs.

\section{Observations \label{sec:observations}}

\noindent{\bf OVRO Millimeter Array.}
Continuum observations in the 3-mm band were made with the Owens
Valley Radio Observatory (OVRO) six element array on 1999 January
28.39 UT. We obtained about 5 hours of on-source integration time in
good 3-mm weather. 
Two 1-GHz bandwidth continuum channels with central frequencies of
98.48 GHz and 101.48 GHz were observed simultaneously in the upper and
lower side bands. The quasar 3C345 was used for phase calibration
and 3C273 for absolute flux calibration; the observing procedure
was otherwise similar to that described in Shepherd et al. (1998).

\noindent{\bf OVRO 40-m Telescope.}
The position of the optical transient was observed with the newly
commissioned 30-GHz receiver on the 40-m OVRO telescope.  The receiver
has two beams, each with a full-width at half maximum (FWHM) of 1.5
arcminute, separated by 7.6 arcminute on the sky.  The two beams were
switched at 125 Hz and flux calibration was achieved by observations
of 3C286.  Pointing was monitored by frequent (20 minutes) observations
of J1419+383 and J1613+342.

\noindent{\bf JCMT}
Observations in the 1.35-mm and 850-$\mu$m bands were made at the
James Clerk Maxwell Telescope (JCMT)\footnotemark\footnotetext{The
  JCMT is operated by The Joint Astronomy Centre on behalf of the
  Particle Physics and Astronomy Research Council of the United
  Kingdom, the Netherlands Organization for Scientific Research, and
  the National Research Council of Canada.} using the Submillimeter
Common-User Bolometer Array (SCUBA; Holland et al.~1998).  The
atmospheric opacity in the 850 $\mu$m band was determined from skydip
observations and that in the 1.35 mm band was inferred from skydips
made at 225 GHz using a tipping radiometer.  The short term
variability of the atmospheric emission was estimated from the outer
ring of pixels and subtracted from the central (photometric) pixel.
Frequent observations of a blazar established that the pointing of the
detector was better than 2 arcsecond. Except for the observations on
January 24, the rest of the data were taken under excellent sky 
conditions.

\noindent{\bf VLA.} 
All observations were obtained in two independent 50-MHz channels
centered on 8.43 GHz and 8.48 GHz.  The interferometer phase was
calibrated using the nearby point source J1545+478. The flux scale was
tied to the source 3C286. Data calibration and imaging were carried
out with the AIPS software package following standard practice.  A log
of the observations and a summary of the results can be found in
Table~\ref{tab:Table-VLA}.

\section{Results \label{sec:results}}

\noindent{\bf OVRO Millimeter Array.}
The source was not detected on January 28. The final image has an
effective central frequency of 99.98 GHz, a synthesized beam of
$1.25'' \times 0.83''$ (FWHM) at position angle $88.1^\circ$, and an rms
noise level $\sim 1.1$ mJy/beam.  At the position of the OT we find a
peak brightness of $-1.44\,$mJy/beam.

\noindent{\bf OVRO 40-m.} 
Each observing run was about 2 hours long and the epochs of the four
runs were (UT) January 24.5, February 5.5, February 6.5 and February
7.5.  Bad weather prevented us from reaching thermal limits by almost
a factor of 10.  Averaging the four datasets we can place  3-$\sigma$
upper limits of 13 mJy (Band A, 26--28 GHz), 10 mJy (Band B, 28--30
GHz), 12 mJy (Band C, 30--32 GHz) and 17 mJy (Band D, 32--34 GHz).

\noindent{\bf JCMT.}
As can be seen from Table~\ref{tab:Table-VLA} the source was not
detected at any epoch save perhaps that of February 4.81 UT. On this
epoch a source is seen with a signal-to-noise ratio (SNR) of 3.3,
nominally a significant detection.  However, single dish observations,
unlike interferometric observations, are susceptible to several
systematic errors. Our confidence in this detection would be
strengthened if the if the source was well detected at least at one
other epoch.

\noindent{\bf VLA.} 
A strong source was detected in only one observation (January
24; see Table~\ref{tab:Table-VLA}).  This unresolved source,
hereafter VLA J152530.3+444559,  is located at (epoch J2000)
$\alpha$\ =\ $15^h25^m30.31^s$ ($\pm{0.01^s}$) $\delta$\
=\ $+44^\circ45^\prime59.24^{\prime\prime}$ ($\pm{0.15}$).
The coordinates of this source are in excellent agreement with the
position of the optical transient (Kulkarni et al. 1999).  As can be
seen from Figure~\ref{fig:lightcurves} the source rapidly fades and is
only marginally detected (at the 2-$\sigma$ level) over the next few
weeks. Other than the detection on January 24th, there was no reliable
detection at any other frequency or at any other epoch including other
radio efforts (Galama et al. 1999).  To our knowledge, this phenomenon --
a short-lived radio emission or flare -- is a new phenomenon in radio
afterglow studies.

Given our claim of the discovery of a new phenomenon, it is important for
us to demonstrate confidence in the existence of the source
VLA J152530.3+444559.
To this end we make the following
observations.  (1) The source is seen in many different subsets of
the data of January 24.  (2) The source is seen in both (circular) hands
of polarization (the fractional circular polarization is less than 37\%
at the 99.9\% confidence level).  (3) The source is separately detected
in each of our 50-MHz wide channels.  (4) Finally, within errors the source
has
the expected point-spread function. In view of these tests, we conclude
that the source VLA J152530.3+444559 is indeed a real source.

Around the epoch of the radio discovery, both optical (Kulkarni et al.
1999; Castro-Tirado et al. 1999, Galama et al. 1999) and X-ray afterglow
(Heise et al. 1999) were strong and fading.  In order to compare the
radio observations with those at other wavelengths we parameterize the
temporal behavior of the 8.46-GHz flux by a two-component power-law
model: flux at time $t$, $f(t)= f_*(t/t_*)^{\alpha_{\rm r,d}}$.  Here,
the time $t$ is measured with reference to the GRB event and $t_*=1.24$
d is the epoch of the VLA detection of the radio emission; $f_*$ is the
8.46-GHz flux at that epoch and the subscript ``r'' (``d'') stands for
``rise'' and refers to  $t<t_*$ (``decay'' with  $t>t_*$).
Between January 23.63 and 24.65, we find the flux rose with  $\alpha_{\rm
r}=0.83$ (mean estimator).  From a Monte Carlo analysis  we find $0.45 <
\alpha_{\rm r} < 1.85$ (95\%  confidence level).  For $t\geq t_*$ we find
$f_*=250\,\mu$Jy and $\alpha_{\rm d}=-1.3$ (14 d.o.f., $\chi^2=11.6$).
At the 95\%-confidence level, $\alpha_{\rm d} < -0.9$ and is $<-0.8$
at the 99.9\%-confidence level. Clearly the source faded and we can,
with great certainty, rule out light curves (for $t>t_*$) with rising,
constant or even slowly decaying emission.

\section{Origin of the Radio Flare \label{sec:origin}}

The simplest possibility is that the radio flare was amplification due
to interstellar scattering and scintillation (ISS).  However, if this
was a typical radio afterglow we expect the afterglow to grow stronger
with time along with additional opportunities for scintillation.
As can be seen from Figure~\ref{fig:lightcurves} the source
was never reliably detected after January 24 thereby rendering the
ISS hypothesis unlikely.  

Sari \&\ Piran (1999) suggest that both the prompt optical emission
(Akerlof et al. 1999) and our radio flare arise from the reverse
shock.  In this model, the particles in the reverse shock cool
adiabatically while there is a continual supply of newly energized
particles in the forward shock.  Consequently, with the passage of
time, the emission from the reverse shock shifts rapidly to lower
frequencies.  At early times, self-absorption suppresses the radio
flux and at late times adiabatic cooling results in weak emission. Due
to these competing effects the emission from reverse shock is seen
briefly -- a radio flare.  In contrast, Galama et al. (1999) have
proposed that the radio flare arises in the forward shock i.e. the
usual afterglow emission.  We now investigate this proposal in some
detail.

At a given time, 
the broad-band afterglow spectrum can be described by three power
laws and three characteristic frequencies
(e.g. Sari, Piran \&\ Narayan 1998):  ({\it i}) $f_\nu\propto
\nu^{-(p-1)/2}$ at high frequencies, $\nu>\nu_m$; ({\it ii})
$f_\nu\propto \nu^{2}$ at low (radio) frequencies, $\nu<\nu_a$ with
({\it iii}) $f_\nu\propto \nu^{1/3}$ bridging the two regimes.  Here,
$p$ is the power law index of the shocked particles in the forward
shock and is usually about 2.5.  $\nu_m$ is the frequency at which the
broad-band spectrum peaks (peak flux is indicated by $f_m$) and $\nu_a$
is the frequency at which the synchrotron optical depth is unity.

As noted by Galama et al. (1999), the multi-wavelength data at the
epoch of the VLA 8.46-GHz detection (epoch, $t_*$) can be fitted to an
afterglow model with $\nu_m\sim 500$ GHz and $\nu_a=7.5$ GHz.  In this
model, one expects the 8.46-GHz flux to initially rise $\propto
t^{1/2}$ and decay when $\nu_m$ falls below $\nu_a$. This is expected
to take place at epoch $t_*(500\,{\rm GHz}/\nu_*)^{2/3}\sim 19t_*$;
here, $\nu_*=8.46$ GHz.  This expectation is in clear contradiction
with our measurements (\Sref{sec:results}). 

The rapid fall in the radio data thus requires that $\nu_m$ be
comparable and preferably below $\nu_*$.  Galama et al. (1999) noting
this requirement suggested a model with $\nu_m\sim \nu_a \sim 30$ GHz
at epoch $t_*$.  In Figure~\ref{fig:bbspectrum} we display the
spectrum with $\nu_a=\nu_m=39$ GHz at epoch $t_*$; hereafter, we refer
to this as model ``${\rm F}_{\rm G}$''. The specific value of $\nu_m$
was chosen so the spectrum fits both the optical $r$-band flux and the
VLA 8.46 GHz detection. Model ${\rm F}_{\rm G}$ violates the 
15 GHz and 86 GHz upper limits. Furthermore, in this model,
we
expect the radio flux to start decaying at epoch
$t_*(\nu_m/\nu_*)^{2/3}$ or 3.5 d -- in violation of the VLA
observations.  The true situation is even worse. When $\nu_m<\nu_{a}$,
the latter will not remain time invariant but evolves as
$\nu_{a}\propto t^{-(3p+2)/2(p+4)}\cong t^{-0.73}$.  This should cause
the flux to increase faster as $t^{5/4}$.  Only when $\nu_{a}$ falls
below $\nu_*$ will the 8.46-GHz emission start decaying and this will
not happen until $8.2t_*\cong 10$ d.  The above argument used
asymptotic limits of the broad-band curve.  In
Figure~\ref{fig:lightcurves} we display the evolution of the models
with spectra shown in Figure~\ref{fig:bbspectrum} using the usual
prescription (e.g. Sari et al. 1998). The refined light curves, while
smoother, do not agree with the observations.

To conclude, in the framework of standard afterglow models (by which
we mean models which assume a power law shocked particle spectrum and
constant fraction of energy in electrons and magnetic fields, relative
to the thermal energy of the shocked particles, no choice of $\nu_m$
and $\nu_a$ (including the case when $\nu_m<<\nu_a$) can explain the
rapid decay of the radio flux {\it together} with the observed optical
afterglow light curve. Our calculations are thus in disagreement with
the model proposed by Galama et al. (1999); essentially the same
difficulty exists with the models used by Wang, Dai \&\ Lu (1999) and
Dai \&\ Lu (1999).  Shi \&\ Gyuk (1999) overcome this difficulty by
appealing to special geometry (rapid interaction of the blast wave
with a cloudlet).  Apart from invoking tremendous energy in the burst
($10^{55}$ erg) their model is unable to account for the smoothly
decaying optical afterglow light curve.

Galama et al. (1999) speculate that the broadening of the characteristic
frequencies ($\nu_a,\nu_m,\nu_c$) may lead to a rapid decay of the radio
flux. In our opinion, broadening the characteristic frequencies will
smooth the predicted light curve but this does not help address
the fundamental problem of rapid decay. 
In the framework of the forward shock model, the radio flare
can be accounted only if the characteristic frequencies can 
be evolved much faster than that given
by standard dynamics.
In contrast, the reverse shock
model provides a natural and consistent explanation for the radio flare.

\section{Radio Emission From the Forward Shock \label{sec:forward}}

Accepting the conclusion that the radio flare arises in the reverse
shock we now discuss the radio emission at late times, $t\gg t_*$.  At
such late times, we expect some radio emission from the forward shock
but with conventional properties, i.e., radio flux slowly rising to a
maximum value ($f_m$) at time $t_m\sim$ weeks, and then decaying in
much the same way as the optical afterglow.  This is an inevitable
consequence of the observed optical afterglow emission.

\begin{figure*}
\centerline{\hbox{\psfig{figure=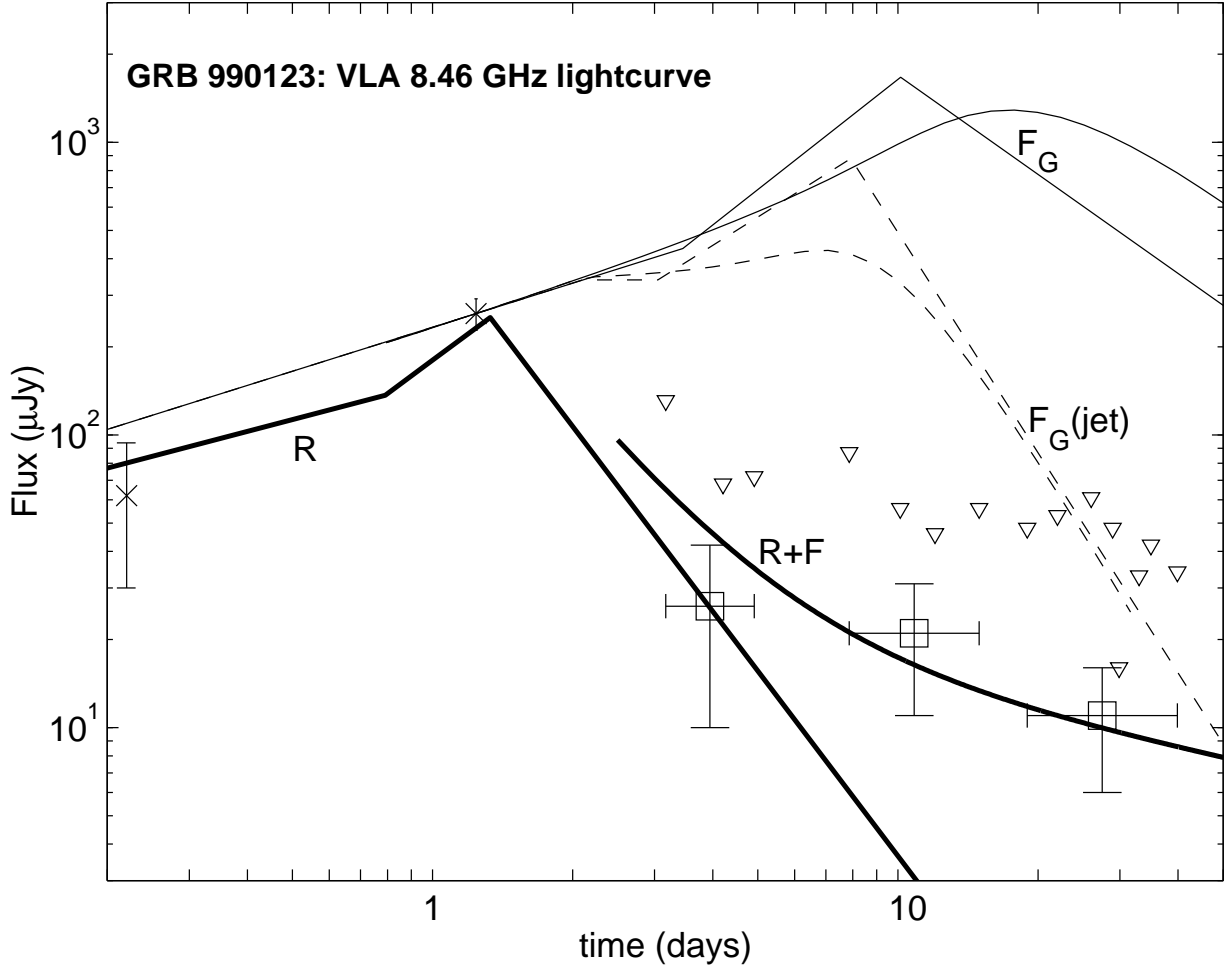,width=6.5in}}}
\caption[]{The observed and model lightcurves at 8.46\,GHz. Triangles
  are upper limits being defined here as the measured flux $+2\sigma$.
  The weighted averages over ranges of days are shown by boxes; the
  vertical size is $\pm \sigma$.  (A) The heavy curves are the
  expected light curves from our model. The left curve is the predicted
  emission from the reverse shock model of Sari \&\ Piran (1999); hereafter
  model ``R''.
  The right curve is the sum of the
  radio emission from the reverse and the forward shock advocated in
  the text (model ``F'', see also Figure~\ref{fig:bbspectrum}) 
  but with a jet geometry.  
  We assume that the jet begins to spread
  at epoch 2.1 d (see \S\ref{sec:forward}). 
  (B) The light solid lines are the light
  curves computed for a conventional spherical forward shock model with
  parameters favored by Galama et al. (1999), $\nu_{a}\sim \nu_{m}\sim
  30$ GHz (model ``${\rm F}_{\rm G}$'', see \S\ref{sec:origin}).  The
  smooth curve is the curve obtained from exact calculations of this
  model whereas the other curve is built up from asymptotic formulae.
  (C) The curves
  with dashed lines refer to the ${\rm F}_{\rm G}$ model but with a
  jet geometry similar to that discussed in (A) above.
\label{fig:lightcurves}} \end{figure*}

\begin{figure*}
\centerline{\hbox{\psfig{figure=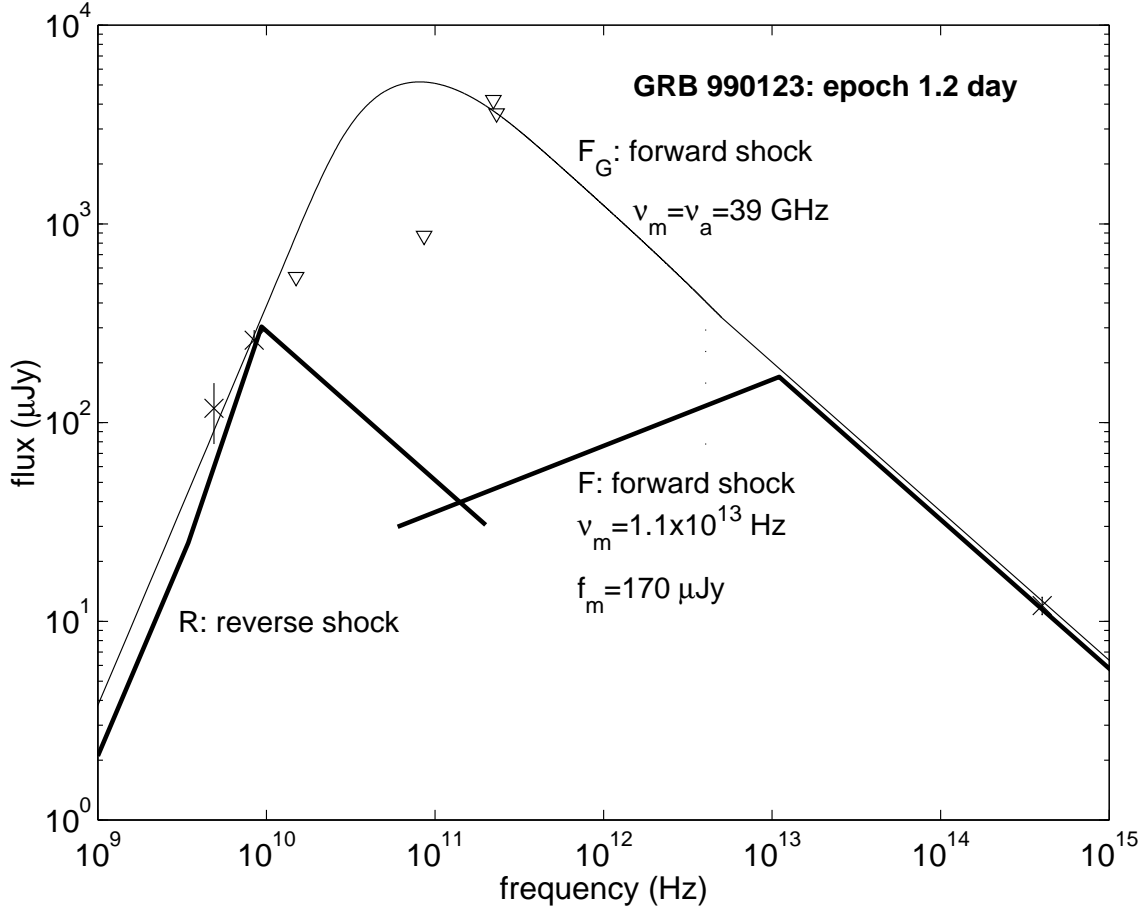,width=6.0in}}}
\caption[]{ The observed broad-band spectrum close to the epoch of Jan
  24.65. The radio data are 4.84 GHz (Galama et. al. 1999, epoch of
  Jan 24.46), 8.46 GHz (this paper, Jan 24.65), 15 GHz (Galama et al.
  1999, Jan 25.26), 86\,GHz and 232\,GHz (Jan 25.18; Galama et al.
  1999) and 222\,GHz (this paper, Jan 24.68). Triangles are $3\sigma$
  upper limits.  The optical point is obtained from the power law
  interpolation of the $r$-band light curve (index $-1.1$) of Kulkarni
  et al. (1999).  The light solid curve labelled ``${\rm F}_{\rm G}$"
  is the expected forward shock contribution from model ``${\rm
    F}_{\rm G}$'' (see caption to Figure~\ref{fig:lightcurves}).  The
  spectral slope at high frequency (optical) is set to $-0.75$,
  consistent with the observed rate of decay of the optical afterglow
  (see Kulkarni et al. 1999).  See the text for details on the
  computation of the curve.  The heavy solid curve ``R" is the
  predicted radio emission from reverse shock model (Sari \&\ Piran
  1999).  The heavy solid curve ``F" is the radio afterglow emission
  from the forward shock; the parameters for this model are displayed
  in the Figure.
\label{fig:bbspectrum}}
\end{figure*}

The highest measured afterglow flux (excluding the ROTSE observations
that we associate with the reverse shock emission) is the Gunn
$r$-band flux, $f_r=100\,\mu$Jy at epoch $t_1\sim 3.7$ hr (Kulkarni et
al. 1999).  Clearly, $f_m \geq f_r$ and $t_r$, the epoch at which the
$r$ band flux peaks is $\leq t_1$.  Thus we expect a peak flux $\geq
f_r$ in the 8.46-GHz band and this peaks at epoch
$t_r(\nu_r/\nu_*)^{2/3}\leq 210$ d; here, $\nu_r$ is the center
frequency of the $r$-band, $\sim 4\times 10^{14}$ Hz.  From this it
follows that we should have seen at least $44\,\mu$Jy in $\leq 40$ d
after the burst.  As can be seen from Figure~\ref{fig:lightcurves} and
Table~\ref{tab:Table-VLA} (also \Sref{sec:results}) our observations
are sensitive enough to exclude such strong emission. Radio emission
can be suppressed by invoking $\nu_{a}>\nu_*$. However, this condition
would also suppress the radio emission from the reverse shock and thus
be unable to account for the radio flare (see \Sref{sec:origin}).

Kulkarni et al. (1999) have interpreted the break in their $r$-band
afterglow emission as due to a jet.  The lack of significant radio
emission is in excellent agreement with the jet hypothesis.  Let $t_b$
be the time at which the jet's lateral dimensions become visible to
the observer and/or the spreading of the jet begins.  If $\nu<\nu_a$
then the radio flux is $\propto t^0$ but $\propto t^{-1/3}$, otherwise
(Sari, Piran \&\ Halpern 1999).  We do note that these scalings for a
spreading jet are different from those derived by Rhoads (1998) who
assumed $\nu_m\ll\nu_{a}$.

From the $r$-band observations we obtain $t_b=2.1$ d (Kulkarni et al.
1999).  Up to epoch $t_b$, the radio flux increases as $t^{1/2}$ and
the expected radio flux at epoch $t_b$ is $10 (f_m/100\,\mu{\rm
  Jy})^{13/9} \mu$Jy.  Limiting to $t>t_b$ we fit a $t^{-1/3}$ power
law to observed fluxes after subtracting the contribution from the
reverse shock (see Figure~\ref{fig:lightcurves}; also Sari \&\ Piran
1999 for details of the reverse shock model).  The best fit model
yields $22(t/t_b)^{-1/3}\,\mu$Jy from which we infer $f_m=170\,\mu$Jy.
At the 95\% confidence level, the radio emission is constrained to be
less than $40(t/t_b)^{-1/3}\,\mu$Jy which implies $f_m<260\,\mu$Jy.
An additional and independent constraint comes from our late time (May
1999) observations (these observations were not used in the fits shown
in Figure~\ref{fig:lightcurves}).  At the 95\% level, our model
predicts $11\,\mu$Jy which should be compared to the observed mean
flux of $7\pm 7\,\mu$Jy (see Table~\ref{tab:Table-VLA}).

Combining the $f_m$ inferred from the radio observations with the
$r$-band power law decay (flux $\propto t^{-1.1}$), we find $t_r>1.6$ hr.
Earlier in this section we noted some constraints on $f_r$ and $t_r$.
Taken together, we find $100 < f_m < 260\,\mu$Jy and $1.6<t_r<3.7$ hr.
Several authors (e.g. Galama et al. 1999, Castro-Tirado et al. 1999)
have proposed that the ROTSE light curve smoothly blends into the optical
afterglow emission. This hypothesis yields $f_m\sim 5$ mJy, certainly
inconsistent with our result above. In our opinion, the observational
basis for this assertion is questionable given the differing slopes
of the ROTSE light curve and the optical afterglow light curve.

We end by noting that early time 
radio observations could potentially offer a unique insight into the physics
of the reverse shock (Sari \&\ Piran 1999, M\'esz\'aros \&\
Rees 1999). Motivated thus, we have re-examined archival data on other
bursts and found evidence for a radio flare in GRB 970828.  The specific
incidence of radio flares of 1:8  appears to be significantly larger
than that of prompt optical emission obtained by ROTSE or LOTIS.

\acknowledgements

DAF thanks R. Hjellming and Marc Verheijen for generously giving up
their VLA time so that a fast response could be made. RS is supported
by a Fairchild Fellowship.


\def\pni{\par\noindent}
\bigskip
\centerline{\bf REFERENCES}

\pni
        Akerlof, C. W. 1999
        Nature, 398, 400


\pni    
        Castro-Tirado, A. J. et al. 1999,
        Science, 283, 2069

\pni
        Dai, Z. G. \&\ Lu, T. 1999,
        astro-ph/9904025


\pni
        Frail, D. A. et al. 1997b,
        Nature, 389, 261





\pni    Frail, D. A., Waxman, E. \&\ Kulkarni, S. R. 1999,
        in prep

\pni    
        Fruchter, A. S. et al. 1999,
        ApJ, 519, L13

\pni    
        Galama, T. et al. 1999,
        Nature, 398, 394
        

\pni
        Heise, J. et al. 1999, 
        Nature, submitted

\pni
        Hjorth, J. et al.
        Science, 283, 2073

\pni 
        Holland, W.S., Robson, E.I., Gear, W. K., Cunningham, C.R., Lightfoot,
        J. F., Tim Jenness, T., Ivison, R. J., Stevens, J. A., Ade, P. A. R.,
        \&\ Griffin, M. J.  1998,  
        MNRAS, 303, 659


\pni
        Kulkarni, S. R. et al. 1999,
        Nature, 398, 389


\pni    
        M\'esz\'aros, P. \&\ Rees, M. R. 1999, astro-ph 9902367.




\pni
        Rhoads, J. E. 1999,
        preprint


\pni
        Sari, R., Piran T. 1999,
        ApJ, 517, L109

\pni
        Sari, R., Piran T. \& Narayan, R. 1998, ApJ, 497, L17

\pni
        Sari, R., Piran T. \& Halpern, J. 1999, 
        ApJ, 519, L17

\pni
        Shepherd, D. S., Frail, D. A., Kulkarni, S. R. \&\
        Metzger, M. R. 1998,
        ApJ, 497, 859

\pni
        Shi, X. \&\ Gyuk, G. 1999,
        astro-ph/9903023
\pni
        Taylor, G. B., Frail, D. A., Kulkarni, S. R.,
        Shepherd, D. S., Feroci, M. \&\ Frontera, F. 1998,
        ApJ, 502, L11

\pni
        Wang, X. Y., Dai, Z. G. \&\ Lu, T. 1999,
        astro-ph/9906062


        

\begin{deluxetable}{lrrrc}
\tabcolsep0in\footnotesize
\tablewidth{\hsize}
\tablecaption{Radio Observations of \grb\ \label{tab:Table-VLA}}
\tablehead {
\colhead {Epoch (UT)}      &
\colhead {Telescope}    &
\colhead {$\tau$ (min)} &
\colhead {Band (mm)}    &
\colhead {S$\pm\sigma$ ($\mu$Jy)}
}
\startdata
1999 Jan  23.63 & VLA  & 15  & 36     &  62$\pm$32       \nl
1999 Jan  24.65 & VLA  & 15  & 36     & 260$\pm$32       \nl
1999 Jan. 24.68 & JCMT & 47  & 1.35   & $-4640\pm2410$   \nl
1999 Jan  26.58 & VLA  & 10  & 36     &  53$\pm$39       \nl
1999 Jan  27.62 & VLA  & 26  & 36     &  18$\pm$25       \nl
1999 Jan. 27.82 & JCMT & 37  & 1.35   & $ 0.00\pm1700$   \nl
1999 Jan. 27.88 & JCMT & 75  & 0.85   & $-3100\pm1100$   \nl
1999 Jan  28.32 & VLA  & 37  & 36     &   22$\pm$25      \nl
1999 Jan. 29.88 & JCMT & 60  & 0.85   & $ 630\pm1700$   \nl
1999 Jan  31.27 & VLA  & 76  & 36     &   49$\pm$19      \nl
1999 Feb  02.52 & VLA  & 25  & 36     &   4$\pm$26       \nl
1999 Feb  04.43 & VLA  & 78  & 36     &  14$\pm$16       \nl
1999 Feb. 04.81 & JCMT & 40  & 0.85   & $ 4790\pm1440$   \nl
1999 Feb. 05.83 & JCMT & 90  & 0.85   & $ 700\pm1200$   \nl
1999 Feb. 06.84 & JCMT & 45  & 0.85   & $1950\pm1820$   \nl
1999 Feb  07.36 & VLA  & 48  & 36     &  10$\pm$23       \nl
1999 Feb. 07.83 & JCMT & 15  & 0.85   & $ 4100\pm2900$   \nl
1999 Feb. 07.85 & JCMT & 30  & 1.35   & $-800\pm2500$   \nl
1999 Feb  11.37 & VLA  & 109 & 36     &  16$\pm$16       \nl
1999 Feb  14.44 & VLA  &  44 & 36     &  17$\pm$18       \nl
1999 Feb  18.43 & VLA  &  31 & 36     &  27$\pm$17       \nl
1999 Feb  21.34 & VLA  & 136 & 36     &  24$\pm$12       \nl
1999 Feb  22.29 & VLA  & 145 & 36     &  $-6\pm$11       \nl
1999 Feb  25.37 & VLA  &  46 & 36     &  $-3\pm$18       \nl
1999 Feb  27.37 & VLA  &  86 & 36     &  14$\pm$14       \nl
1999 Mar  04.27 & VLA  & 132 & 36     &   8$\pm$13       \nl
1999 May  06.29 & VLA  & 110 & 36     &   6$\pm$10       \nl  
1999 May  07.41 & VLA  & 170 & 36     &   9$\pm$10       \nl  
\enddata
\tablecomments{
The columns are (left to right),
(1) Start of observations (UT date),
(2) Telescope 
(3) Duration of observations, 
(4) Band of observations,
(5) Mean flux density and rms ($\mu$Jy) at the 
position of the radio transient.
\hb
}
\end{deluxetable}


\end{document}